\documentclass[aps,prx,twocolumn,groupedaddress,amsmath,amssymb]{revtex4}

\usepackage{graphicx}
\usepackage{color}
\usepackage{comment}
\usepackage{bm}
\usepackage{latexsym}

\usepackage[symbol]{footmisc}
\usepackage{hyperref}

\newcommand{\beginsupplement}{%
        \setcounter{table}{0}
        \renewcommand{\thetable}{S\arabic{table}}%
        \setcounter{figure}{0}
        \renewcommand{\thefigure}{S\arabic{figure}}%
     }

\usepackage[T1]{fontenc}
\usepackage[latin9]{inputenc}
\usepackage{esint}

\begin{document}

\title{Controlling Optical Beam Thermalization via Band-Gap Engineering}
\author{Cheng Shi}
\affiliation{Wave Transport in Complex Systems Lab, Physics Department, Wesleyan University, Middletown CT-06459, USA}
\author{Tsampikos Kottos}
\thanks{tkottos@wesleyan.edu}
\affiliation{Wave Transport in Complex Systems Lab, Physics Department, Wesleyan University, Middletown CT-06459, USA}
\author{Boris Shapiro}
\affiliation{Technion-Israel Institute of Technology, Technion City, Haifa 32000, Israel}

\begin{abstract}
We establish dispersion engineering rules that allow us to control the thermalization process and the thermal state of an 
initial beam propagating in a multimode nonlinear photonic circuit. To this end, we have implemented a kinetic equation 
(KE) approach in systems whose Bloch dispersion relation exhibits bands and gaps. When the ratio between the gap-width 
to the band-width is larger than a critical value, the KE has stationary solutions which differ from the standard Rayleigh-Jeans 
(RJ) distribution. The theory also predicts the relaxation times above which such non-conventional thermal states occur. We 
have tested the validity of our results for the prototype SSH model whose connectivity between the composite elements 
allows to control the band-gap structure. These spectral engineering rules can be extended to more complex photonic networks 
that lack periodicity but their spectra consist of groups of modes that are separated by spectral gaps. 
\end{abstract}

\maketitle

\section{Introduction }

The understanding of light propagation in nonlinear multimode settings has both fundamental and technological ramifications. Optical 
phase transitions \cite{SF20,KSVW10,S12,RFKS20}, beam self-cleaning \cite{KTSFBMWC17,LWCW16,N19} (a similar phenomenon 
occurs also for vibrational polar modes \cite{ZAS19}), spatio-temporal mode 
locking \cite{WCW17}, and multimode solitons \cite{WCW15} are some of the novel phenomena that these studies have recently revealed. 
At the same time, the development of predictive tools of the nonlinear beam dynamics will be addressing urging technological needs 
associated with the looming information ``capacity crunch'' in fiber-optical communication systems \cite{HK13,RFN13} and the quest 
for new platforms of high-power light sources \cite{WCW17}. Yet, the evaluation of energy redistribution among the various modes due 
to the nonlinear interaction is a daunting task. Typically, it is addressed via brute-force computations -- a formidable process which involves 
a mode-by-mode analysis \cite{A12}. For example, the light propagation in multimode nonlinear fibers requires knowledge of the group 
velocity of each mode, their propagation constant, the self-phase modulation coefficients (which are proportional to $M$ for $M$-modes), 
the cross-phase modulation coefficients (proportional to $M^2$), the four wave mixing coefficients (proportional to $M^3$), etc.

This kind of a brute force calculation would give the detail picture of the time evolution of energy distribution over the (linear) modes of 
the system. Often, however, one is interested only in the long-time limit of the distribution. The latter follows from the general principles 
of statistical mechanics which imply that a collection of weakly interacting modes reaches an equilibrium distribution given by the Rayleigh-
Jeans (RJ) formula \cite{Petal14,DNPZ92}. When the interacting optical modes reach an internal equilibrium, one can define various 
thermodynamic functions and develop an ``optical thermodynamics'' \cite{WHC19}, in complete analogy with the standard theory for 
systems in thermal equilibrium \cite{LLv1}. We will often refer to this internal equilibrium in the mode space as ``thermal equilibrium'' 
although the corresponding temperature is only an effective one which has nothing to do with the actual temperature of the physical 
environment.

The assertion of the RJ distribution, in the long-time limit, has been supported by deriving and studying an approximate kinetic equation 
for the intensity distribution of the modes in case of Kerr type non-linearities \cite{Petal14,N11}. This equation exhibits a stationary solution 
of the RJ form. The same conclusion has been reached using the optical thermodynamic approach \cite{WHC19,PWJMC19,WJPKC20}. 
Importantly, this framework, immediately implies that the formation of a RJ equilibrium distribution is independent of the specific nature of 
the nonlinear interaction (e.g. Kerr or saturable or thermal nonlinearities) as long as the latter is weak \cite{WHC19,MWJC20,RFKS20}. 
The RJ distribution has been recently confirmed by a direct measurement of the thermal state in a highly-multimode optical fiber 
\cite{PSWWCW21}. Other indirect measurements of RJ have provided additional confidence on the validity of these theoretical predictions 
\cite{BFKGRMP20}.

In contrast to the optical thermodynamic approach that takes thermalization as granted, the kinetic equation provides also an understanding 
of the thermalization process. Unfortunately, its mathematical complexity has limited its implementation only to simple 
systems, with a  linear spectrum consisting of a single Bloch band (like the photonic lattice in Fig. \ref{fig1}a). In this case one always ends 
up with a stationary RJ distribution whose effective temperature and chemical potential are uniquely determined from the initial norm and 
energy. The main purpose of the present paper is to demonstrate that in more complex cases, when the linear spectrum of the system 
exhibits bands and gaps, the kinetic equation can have, under specific conditions, stationary solutions with different RJ distributions in 
different bands. These solutions describe states of a partial (or quasi) equilibrium of the system. Our analysis 
underlines the importance of band-gap engineering in the Bloch dispersion relation of a composite photonic circuit and highlights the relaxation 
processes that are responsible for the thermalization of an initial beam towards this (quasi-) stationary {\it two-component RJ} (TCRJ) distribution.
Our theoretical considerations are confirmed via detailed numerical simulations with a variety of weakly nonlinear multimode systems.

The structure of the paper is as follows. First we present the theoretical model that describes the beam propagation in nonlinear multimode 
(multicore) fibers and photonic networks of coupled resonators. The associated kinetic equation and its stationary RJ solution is briefly 
reviewed as well. Then in Sec. \ref{GRJ} we analyze the conditions under which a thermal equilibrium states follows the two-component 
RJ distribution. This analysis allows us to establish a set of spectral engineering rules that determine the nature of the thermal state. In 
the next Sec. \ref{manage} we provide examples of thermal equilibrium management in composite photonic networks with underlying 
periodicity. A prominent example that we analyze in detail is the Su-Schrieffer-Heeger (SSH) photonic lattice for which we also derive 
theoretical expressions for the relaxation times towards the thermal equilibrium. In Sec. \ref{random} we extend the spectral engineering 
approach to random photonic lattices. Our conclusions and outlook are summarized at the last section \ref{conclusions}.

\section{Modeling of Beam Dynamics in Photonic Networks} \label{modeling}
\subsection{Dynamical Equations}
We model the dynamics of multimode nonlinear photonic networks using a time-dependent 
coupled mode theory (CMT)
\begin{equation}\label{cmt}
    i\frac{d\psi_l}{dt} = -\sum_{j}J_{lj}\psi_{j} + \chi|\psi_{l}|^2\psi_{l},\indent l = 1,\cdots, M,
\end{equation}
where $\psi_l$ is the complex field amplitude at node $l$ while $J_{lj} = J_{jl}^*$ describes the coupling between the nodes $l$ and $j$. 
The last term models a nonlinear light-matter interaction, which is associated with a Kerr effect. The variable $t$ in Eq. (\ref{cmt}) represents 
(a) either time in the case of temporal field dynamics at a network of coupled micro-resonators or (b) the paraxial propagation 
distance of a beam propagating in a multimode fiber or in an array of coupled waveguides. 

\begin{figure*}[ht]
\centering
\includegraphics[width=2\columnwidth]{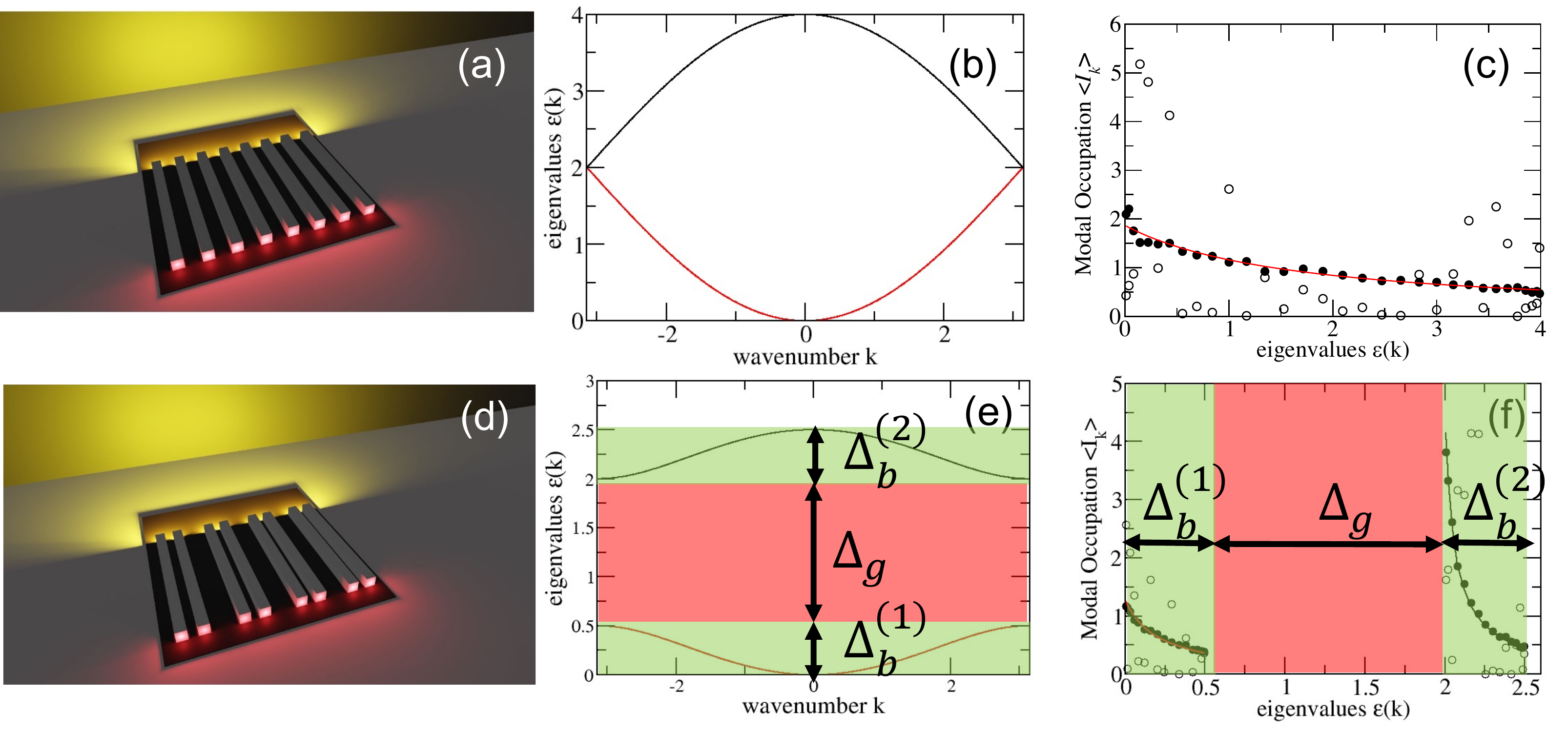}
\caption{
Upper row: (a)A photonic circuit consisting of $M=32$ coupled single-mode nonlinear waveguides. In this numerical example, all waveguides 
have the same propagation 
constant $J_{ll}=2$ (in units of coupling $J_{l,l\pm1}=1$); (b) The dispersion relation $\varepsilon_{\alpha}=\varepsilon(k)$ of the corresponding linear 
system; (c) The modal power occupation $I_k$ of the initial excitation (open circles) and of the thermal state (filled black circles). The solid red line 
is the theoretical prediction describing the standard RJ distribution with $(T,\mu)=(3.058,-3.639)$. Lower row: (d) A photonic circuit consisting of 
$M=32$ coupled nonlinear waveguides forming an SSH structure. The waveguides are arranged in dimers with intra-dimer coupling $v=1$ 
and inter-dimer couplings $w=1/4$ (units of $v$). The propagation constant in each waveguide is $J_{ll}=v+w=1.25$ (units of $v$); (e) The dispersion 
relation of the SSH array; (f) The modal power occupation $I_k$ of the initial excitation (open circles) and of the thermal state (filled black circles). 
The solid red and black lines are the theoretical predictions describing the TCRJ distribution with $(T,\mu_1,\mu_2)=(0.258,-0.201,1.943)$. 
In both examples in Figs. \ref{fig1}c,f, the Kerr nonlinearity strength is $\chi=0.05$. In both (c) and (f), the propagation distance is $t_{\rm max}=10^6$ 
(units of coupling constant). 
}
\label{fig1}
\end{figure*}

The equation of motion (\ref{cmt}) is derivable from the Hamiltonian (internal energy) 
\begin{equation}
\label{CMTH}
{\cal H}(\{\psi_l\})=-\sum_{l,j}J_{lj}\psi_l^*\psi_j+{1\over2}\chi\sum_l|\psi_l|^4\equiv E
\end{equation}
which is a constant of motion. Below we will assume that the Kerr nonlinearity coefficient $\chi$ is small. This assumption will allow us to 
approximate the total internal energy (Hamiltonian) by its linear part i.e. ${\cal H}(\{\psi_l\})\equiv {\cal H}_L+{\cal 
H}_{NL}\approx {\cal H}_L$. Physically, in multimode waveguide settings, the internal energy represents the longitudinal electrodynamic 
momentum flow. Another constant of motion is
\begin{equation}
\label{CMTN}
{\cal N}(\{\psi_l\})=\sum_l|\psi_l|^2\equiv A
\end{equation}
which can be interpreted as the total optical power of the beam.

Equation (\ref{cmt}) describes the field dynamics in the Wannier basis associated with the (localized) modes of the individual nodes 
of the circuit. There is an alternative formulation of this problem which utilizes the supermode basis $\{f_{\alpha}\}, \alpha=1,\cdots,M$ 
of the underlying linear network. In this case $\psi_{l}(t) = \sum_{\alpha}C_{\alpha}(t)f_{\alpha}(l)$, and Eq.~(\ref{cmt}) is re-written 
as:
\begin{equation}
\label{mode_rep}
i\frac{dC_{\alpha}}{dt} = \varepsilon_{\alpha}C_{\alpha} + \chi\sum_{\beta\gamma\delta}\Gamma_{\alpha\beta
\gamma\delta}C^*_{\beta}C_{\gamma}C_{\delta},
\end{equation}
where $\varepsilon_{\alpha}$ is the eigenvalue associated with eigenmode $f_{\alpha}$ of the linear network and
\begin{equation}\label{Gamma}
    \Gamma_{\alpha\beta\gamma\delta} = \sum_{l}f_{\alpha}^*(l)f_{\beta}^*(l)f_{\gamma}(l)f_{\delta}(l)
\end{equation}
describes the interactions associated with a nonlinear mixing between supermodes.

\subsection{Kinetic Equation and Standard Rayleigh-Jeans Distribution}

A well established method that allows to analyze the thermalization process of a light beam propagating under Eqs. (\ref{cmt},
\ref{mode_rep}) is based on the nonequilibrium kinetic equation, which in turn is based on the Random Phase Approximation. 
The approximation is valid because any particular mode is coupled, by nonlinearity, to a great number of other modes (essentially 
to a continuum of modes). This coupling leads to a chaotic mode dynamics and to the phase randomization. The system therefore 
can be described solely in term of the slowly varying mode intensities which obey the kinetic equation (KE) \cite{Petal14,N11}: 
\begin{eqnarray}\label{KE}
\frac{dI_{\alpha}}{dt} &=&\chi^2\sum_{\beta\gamma\delta}'V_{\alpha\beta\gamma\delta}\left(I_{\beta}I_{\gamma}I_{\delta}+
I_{\alpha}I_{\gamma}I_{\delta}-I_{\alpha}I_{\beta}I_{\delta}-I_{\alpha}I_{\beta}I_{\gamma}\right)\nonumber\\
&=&\chi^2 \sum_{\beta\gamma\delta}'V_{\alpha\beta\gamma\delta}I_{\alpha}I_{\beta}
I_{\gamma}I_{\delta}(\frac{1}{I_{\alpha}}+\frac{1}{I_{\beta}}-\frac{1}{I_{\gamma}}-\frac{1}{I_{\delta}}),
\end{eqnarray}
where $I_{\alpha}(t) = |C_{\alpha}(t)|^2$ is the optical power in supermode $\alpha$, $\sum_{\beta\gamma\delta}'$ denotes the 
summation over all the non-diagonal terms (triplets satisfying $(\alpha,\beta)\neq (\gamma,\delta)$ and $(\alpha,\beta)\neq (\delta,
\gamma)$), and 
\begin{equation}
\label{V-element}
    V_{\alpha\beta\gamma\delta} = 4\pi|\Gamma_{\alpha\beta\gamma\delta}|^2\delta(\varepsilon_{\alpha}+
\varepsilon_{\beta}-\varepsilon_{\gamma}-\varepsilon_{\delta}).
\end{equation}
It is straightforward to confirm that a stationary solution of Eq.~(\ref{KE}) is the RJ distribution \cite{Petal14,DNPZ92,AGMDP11}
\begin{equation}\label{RJ}
    {\tilde I}_{\alpha} = \frac{T}{\varepsilon_{\alpha}-\mu}
\end{equation}
where the optical temperature and chemical potential $(T,\mu)$, are obtained from the total power $A$ and the total (linear) energy 
$E$ of the initial beam.

Equation~(\ref{KE}) features a four-mode (i.e. excitations) interaction process which respects the conservation of the total power $A=
\sum_{\alpha}I_{\alpha}$ and total (linear) energy $E\approx\sum_{\alpha}\varepsilon_{\alpha}I_{\alpha}$, and admits the following 
physical interpretation. Generally, all four modes $I_{\alpha,\beta,\gamma,\delta}$ appearing in the first line in Eq. (\ref{KE}) are occupied 
(this is the case in thermal equilibrium), although in principle it can happen that one out of the four modes is initially empty but gets occupied 
after the collision. During the 
interaction process, two of these modes will gain some norm (and energy) while the other two will 
lose the same amount. Specifically, the collision process assumes that two (occupied) modes act on the third (occupied) mode, causing 
(by induced emission) an increase of the norm of that third mode and producing some norm in the fourth mode (which can be occupied 
or being empty before the collision). We stress, that the KE approach adopted in this work is applicable only in the weak nonlinearity 
limit where the concept of linear modes is still valid. In the opposite limit of strong nonlinearities, one needs to adopt an alternative approach 
based on the evaluation of the Gibbs distribution \cite{Kevrekidis}.    

Figure \ref{fig1}a displays the simplest photonic lattice, consisting of an array of coupled single-mode waveguides with nearest neighbor 
coupling constants $J_{l,l\pm1}=v(=1)$ and same resonant frequency $J_0\equiv J_{l,l}(=2)$. The photonic lattice has a linear dispersion relation 
consisting of a single band $\varepsilon(k)=2-2v\cos(k)$ where $k\in[-\pi,\pi]$ is the propagation constant (wavevector). For reasons that will 
become clear below, we have chosen to represent the dispersion relation of such system using a reduced Brillouin zone associated with 
doubling of the unit cell. In this representation the dispersion relation takes the form  $\varepsilon(k)=2\pm 2v\cos(k/2)$, see Fig. \ref{fig1}b. 

The thermalization process of a typical initial state (open circles) has been verified numerically in Fig. \ref{fig1}c. In these simulations we 
have integrated Eq. (\ref{cmt}) up to times ${\cal O}(10^{6})$ (in units of coupling constant) using a high-order three part split symplectic 
integrator scheme \cite{RFKS20}. The method conserved, up to errors ${\cal O}(10^{-8})$, the total internal energy and power of the system. 
The thermal state (filled black circles) has been extracted by performing a time averaging over the last $2\times 10^{5}$ propagation times. 
We have found that such system always thermalizes to the standard RJ distribution (red line) Eq. (\ref{RJ}).

\section{Two-Component Rayleigh-Jeans Distributions} \label{GRJ}

We  now argue that more complicated optical networks, whose spectrum involves bands and gaps, can approach in their thermalization 
process a different distribution which we name two-component Rayleigh-Jeans (TCRJ). More precisely, the kinetic equation for such 
systems has stationary solutions which correspond to different RJ distributions in different bands. An example is shown in Fig. \ref{fig1}d 
where the connectivity $J_{lj}$ of the waveguide array (SSH system--see below) is such that the spectrum of the system consists of two 
bands separated by a gap, see Fig. \ref{fig1}e. It turns out that the TCRJ distributions (see Fig. \ref{fig1}f for the SSH example) are 
possible because, under conditions specified below, the optical power is conserved separately in each individual band.  

Strictly speaking, a TCRJ distribution corresponds to a partial equilibrium (a long-lived state) rather than to a full one. The latter 
is eventually established via processes involving exchange of the modal power between the bands. The main point, however, is that these 
processes (again, under the appropriate conditions, see below) are beyond the KE Eq. (\ref{KE}) and would involve time scales much 
longer than those needed for reaching the TCRJ state. Therefore the latter can be considered as a genuine equilibrium state for all practical 
processes. Our goal below is to establish dispersion engineering rules that will allow us to control the thermalization process, and therefore 
the thermal distribution, of an initial beam.\\

\subsection{Collision Processes and Conservation laws} 

To understand the results shown in Fig. \ref{fig1}f we analyze the KE under the assumption that the spectrum of the system has a band-gap 
structure. For simplicity, we consider a system with one energy gap separating the spectrum into two bands. In this case, there are the following 
``collision'' processes between modes that contribute to the relaxation described by Eq. (\ref{KE}): (I) all four interacting modes are from 
the same band; (II) two of the interacting modes belong to one band while the other two belong to the other band; 
(III) three interacting modes are from the same band, and one mode is from the other band. All other scenarios are energetically impossible. 

Processes (II) and (III) exchange energy between the two bands, while only process (III) exchanges both energy and power. The lack 
of process (III), therefore, enforces the {\it conservation of power in each individual band}, i.e., 
\begin{equation}
\label{local}
A^{(1)} = \sum_{\alpha_1}I_{\alpha_1};\quad A^{(2)} = \sum_{\alpha_2}I_{\alpha_2}
\end{equation} 
where ${\alpha_1}$ and ${\alpha_2}$ refer to the modes associated with the first and the second band respectively. A lack of 
processes (II) and (III) altogether will lead to a total disconnection between the two bands. 

Let us analyze further the conditions under which process (III) is possible. For this purpose, we define the gap-width $\Delta_g$, and the 
band-widths $\Delta_b^{(s)}$ of the lower $(s=1)$ and upper $(s=2)$ bands. We further define the corresponding bottom energy states 
$\varepsilon_{\rm min}^{(1)},\varepsilon_{\rm min}^{(2)}=\varepsilon_{\rm min}^{(1)}+\Delta_b^{(1)}+\Delta_g$, see Fig. \ref{fig1}e for an 
example. Next, we consider a collision process which involves three modes from the lower band and one mode from the upper band. In 
this case, the most extreme collision process (that respects the energy conservation) involves the two highest modes of the lower band, 
interacting to form two excitations, each located at the bottom of the corresponding band. These considerations 
enforce the following inequality $2(\varepsilon_{\rm min}^{(1)}+\Delta_b^{(1)})\geq \varepsilon_{\rm min}^{(1)} +\varepsilon_{\rm min}^{(2)}$, 
which can be simplified to $\Delta_b^{(1)}\geq \Delta_g$. A similar argument applies when the three modes that participate in the collision 
process are associated with the upper band $s=2$ and the fourth is associated with the lower band $s=1$. The corresponding extreme 
process, which still allows the presence of process (III), involves two modes from the bottom of the upper band to collide and form two 
modes that are located at the top of the upper and lower band respectively. Consequently, we have the energetic restriction $2\varepsilon_{
\rm min}^{(2)}\leq (\varepsilon_{\rm min}^{(1)}+\Delta_b^{(1)})+(\varepsilon_{\rm min}^{(2)}+\Delta_b^{(2)})$ which leads to $\Delta_g\leq 
\Delta_b^{(2)}$. We conclude therefore that the absence of process (III) is guaranteed whenever the following inequality is satisfied
\begin{equation}
\label{ab3}
\Delta\equiv{\Delta_g\over\Delta_b}\geq 1; \quad \Delta_b\equiv{\rm max}\left(\Delta_b^{(1)},\Delta_b^{(2)}\right)
\end{equation}
leading to the conservation of the total internal energy, and of the individual band-powers $(E,A^{(1)},A^{(2)})$ in Eq. (\ref{KE}). 
 
Under the broad gap condition Eq. (\ref{ab3}), the KE collapses to the following form for the relaxation process associated with the 
$\alpha_1-$modes (and similarly for the $\alpha_2-$modes)
\begin{widetext}
\begin{equation}
\label{normconservation}
    \frac{dI_{\alpha_1}}{dt} = \chi^2\sum_{\beta_1\gamma_1\delta_1}' V_{\alpha_1\beta_1\gamma_1\delta_1}I_{\alpha_1}
I_{\beta_1}I_{\gamma_1}I_{\delta_1}(\frac{1}{I_{\alpha_1}}+\frac{1}{I_{\beta_1}}-\frac{1}{I_{\gamma_1}}-\frac{1}{I_{\delta_1}})
+
2\chi^2\sum_{\beta_2\gamma_1\delta_2}' V_{\alpha_1\beta_2\gamma_1\delta_2}I_{\alpha_1}I_{\beta_2}I_{\gamma_1}
I_{\delta_2}(\frac{1}{I_{\alpha_1}}+\frac{1}{I_{\beta_2}}-\frac{1}{I_{\gamma_1}}-\frac{1}{I_{\delta_2}}) 
\end{equation}
\end{widetext}
which results from Eq. (\ref{KE}) after excluding from the summation the collision processes (III). The first sum at the r.h.s. of Eq. 
(\ref{normconservation}) describes the intra-band collisions (I) while the second term represents the inter-band processes (II). 

\begin{figure}
\centering
\includegraphics[width=1\columnwidth]{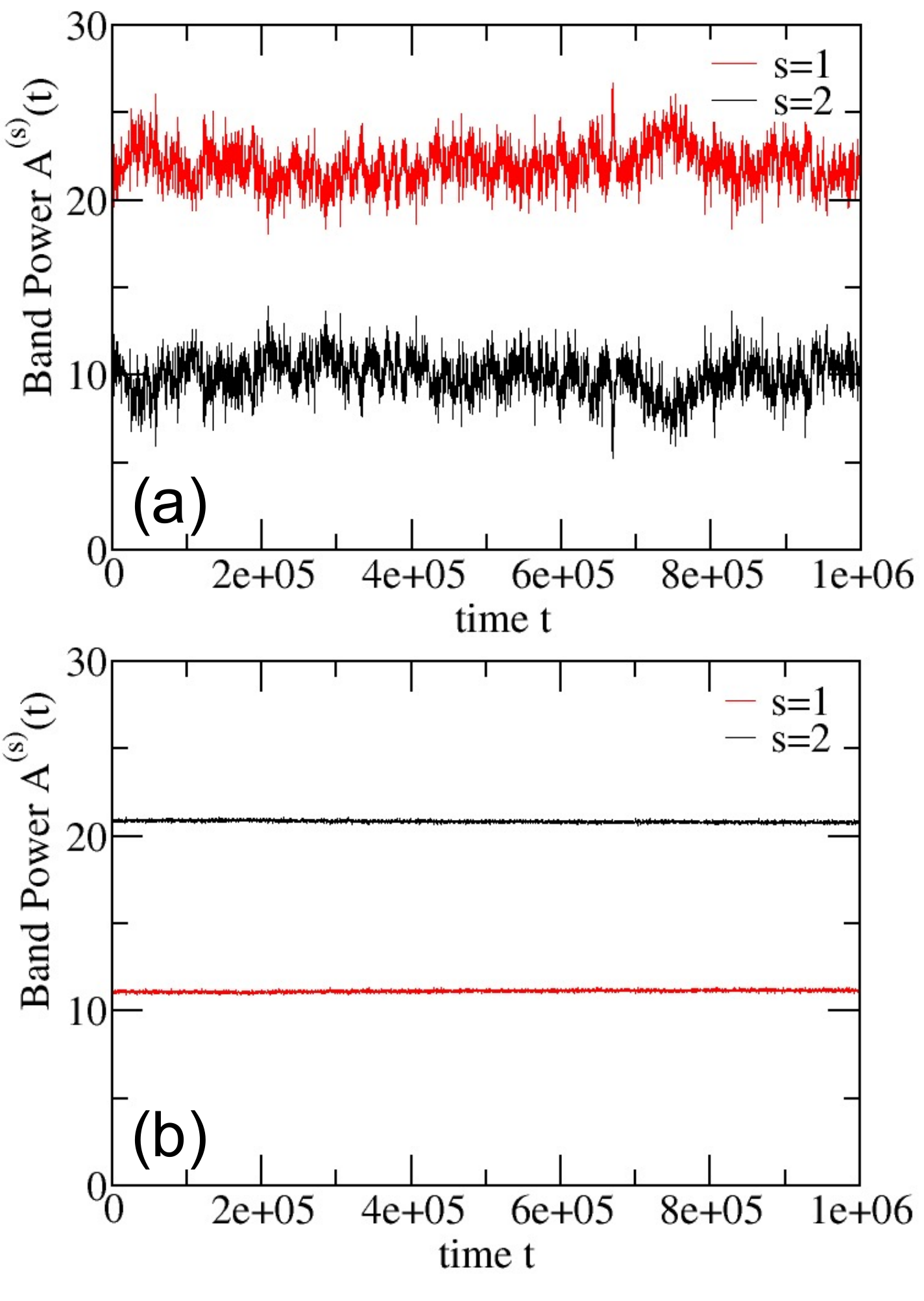}
\caption{
Temporal evolution of the optical power in each band for the two cases of Fig. \ref{fig1}a,d: (a) When the broad-gap condition is 
violated, the band power $A^{(s)}$ oscillates in time; (b) When the broad-gap condition is satisfied, the band power $A^{(s)}$ 
remains constant. In this case $\Delta_g=1.5$, and $\Delta_b=0.5$ corresponding to $\Delta=3$. In these simulations we have 
assumed that the strength of the Kerr nonlinearity is $\chi=0.05$ while the SSH model (see Fig. \ref{fig1}d) has intra-cell coupling 
$v=1$ and inter-cell coupling $w=1/4$
}
\label{fig2}
\end{figure}

Using Eq. (\ref{normconservation}), it is straightforward to show that indeed $\frac{dA^{(s)}(t)}{dt}=0$. A numerical confirmation 
of the consequences of the absence/presence of the broad-gap condition Eq. (\ref{ab3}), is shown in Figs. \ref{fig2}a,b where 
we are reporting our results for $A^{(s)}(t)$, for two representative cases: In the set-up of Fig. \ref{fig1}a the band powers $A^{
(1,2)}(t)$ vary in time (see Fig. \ref{fig2}a). In contrast, in the set-up of Fig. \ref{fig1}d where the connectivity is such that the 
band-spectrum satisfies the broad-gap condition Eq. (\ref{ab3}), both $A^{(1,2)}(t)$ remain constant in time (see Fig. \ref{fig2}b).

\subsection{Two-Component Rayleigh-Jeans Distribution} 

Next we analyze the nature of the thermal state under the broad gap condition. In this case, the distribution governed by Eq. 
(\ref{normconservation}) does not evolve towards the standard RJ distribution i.e., there is no full thermalization. Rather, the 
following TCRJ distribution nullifies the r.h.s. of Eq. (\ref{normconservation})
\begin{equation}
\label{dRJ}
{\tilde I}_{\alpha_1} = \frac{T}{\varepsilon_{\alpha_1}^{(1)}-\mu_1},\indent  {\tilde I}_{\alpha_2} = \frac{T}{\varepsilon_{\alpha_2}^{(2)}
-\mu_2}
\end{equation}
where $\varepsilon_{\alpha_s}^{(s)}$ and ${\tilde I}_{\alpha_s}$ refer to the eigenvalues and modal power associated with the 
$\alpha_{1,2}-$modes of the first $(s=1)$ and second ($s=2)$) band respectively. Equations (\ref{dRJ}) dictate that the initial beam 
relaxes to a uniform optical temperature, but {\it the modes in 
the two bands do not share the same chemical potential}. Consequently the TCRJ distribution is defined by three parameters, $(T,
\mu_1,\mu_2)$. These parameters have an one-to-one correspondence with the conserved quantities $(E,A^{(1)},A^{(2)})$, and can 
be calculated explicitly from the initial beam. Specifically, the conservation of total energy and norm in each band can be expressed 
as: 
\begin{align}
\label{su1}
\sum_{\alpha_1 = 1}^{M_1}\frac{T}{\varepsilon_{\alpha_1}^{(1)}-\mu_1} = A^{(1)},\indent \sum_{\alpha_2 = 1}^{M_2}\frac{T}
{\varepsilon_{\alpha_2}^{(2)}-\mu_2} = A^{(2)}, \nonumber\\
(M_1+M_2) T + \mu_1 A^{(1)} + \mu_2 A^{(2)} = E,
\end{align}
where $M_1$ and $M_2$ refer to the number of modes involved in each band ($M=M_1+M_2$).

Let us re-iterate that the TCRJ distribution might not necessarily be the true equilibrium state of the system. Its existence is 
tightly connected with the validity of conservation laws Eq. (\ref{local}) for $A^{(1)}$ and $A^{(2)}$. The latter is the result of the 
KE Eqs. (\ref{normconservation}), which describes the dynamics generated by  Eq.~(\ref{cmt}) for times up to $t\sim t^{\rm KE}
\propto {\cal O} (1/\chi^4)$. This time constraint signifies the break-down of the second-order approximation for the rate of change 
of $I_{\alpha}$. For even longer times, the effects of higher-order terms in the nonlinear interaction among the modes are going to 
accumulate, and the modal powers will eventually 
converge to the standard RJ distribution. However, the derivation of the broad-gap condition Eq. (\ref{ab3}) for the existence of a 
(quasi-)stationary TCRJ thermal state is extremely relevant to any realistic applications where the paraxial distance (or evolution time) 
is finite.

\section{Examples of Thermalization Management via Dispersion Engineering}\label{manage}

In this section we will be implementing in practice the thermalization management via dispersion engineering. For this purpose, we 
shall analyze the thermal state for a number of photonic circuits corresponding to different connectivity matrices $J_{lj}$. A special 
attention will be given to the thermalization process of the so-called Su-Schrieffer-Heeger (SSH) network which has attracted a 
lot of attention for its topological properties recently. 

It turns out that the SSH model provides a good theoretical framework where the effects of the existence/absence of the broad-gap 
condition Eq. (\ref{ab3}) on the relaxation times can be also analyzed theoretically. These results can further guide our understanding 
of the relaxation process itself.
It is important to point out that photonic systems with separated modes in different groups, which do not exchange power 
with one-another, have been demonstrated also in \cite{WHC19}. However, the design mechanism in Ref. \cite{WHC19} is 
completely 
different from the one discussed here. Specifically, these authors have utilized the polarization degrees of freedom in order 
to enforce a mode-connectivity matrix $V_{\alpha\beta\gamma\delta}$ which suppresses transitions between modes at 
different polarization. In contrast, our design protocol involves spectral engineering methods for the management of the 
thermal equilibrium states.

\subsection{The SSH Model}

A prototype system that demonstrates a transition from a RJ to a TCRJ is the SSH model, see Fig. \ref{fig1}d. 
It consists of $M/2$ unit cells, with each cell containing two single-mode waveguides coupled together with an intra-cell coupling constant 
$v$. The inter-cell coupling is only between neighboring cells and it is described by a coupling constant $w<v$. By setting the
``on-site'' potential (mode) of each waveguide equal to $v+w$, we ensure that the ground-state energy is zero. The corresponding 
coupled mode theory (CMT) Hamiltonian is:
\begin{equation}\label{SSHmodel}
    \mathcal{H} = -\sum_{n,s,n',s'}J_{s-s'}^{n-n'}\psi_{n,s}^{\ast}\psi_{n',s'} + \frac{1}{2}\sum_{n,s}\chi|\psi_{n,s}|^4
\end{equation}
where $s,s'\in\{1,2\}$ and $\psi_{n,s}$ refers to the complex amplitude on the $s$th site in the $n$th unit cell. The connectivity matrix is 
three-diagonal with a diagonal element $J_{0}^0 = -(v+w)$, and off-diagonal elements $J_{\pm 1}^{0} = v$, $J_{\pm 1}^{\mp 1} = w$, 
and $J = 0$ otherwise. 

The eigenmodes of this system can be analytically evaluated and they take the form
\begin{eqnarray}
\label{vectors}
f^{(1)}(k)={1\over \sqrt{2}}\left[\cdots,\left(1,\sqrt{{v+w e^{ik}\over v+w e^{-ik}}}\right),\cdots\right]^T\\
f^{(2)}(k)={1\over \sqrt{2}}\left[\cdots,\left(1,-\sqrt{{v+w e^{ik}\over v+w e^{-ik}}}\right),\cdots\right]^T\nonumber
\end{eqnarray}
where $f^{(s=1,2)}(k)$ is the eigenmode amplitude in the momentum representation i.e. $\psi_{n,s}={1\over 2\pi}\int dk 
e^{ikn}f^{(s)}(k)$ and $k\in[-\pi,\pi]$ is the wavenumber taken over the Brillouin zone. The corresponding eigenvalues are 
\begin{align}\label{SSHspectrum}
    \varepsilon^{(1)}(k) = v + w - \sqrt{v^2 + w^2 + 2vw\cos(k)},\nonumber\\
    \varepsilon^{(2)}(k) = v + w + \sqrt{v^2 + w^2 + 2vw\cos(k)}.
\end{align}
The lower (higher) band $\varepsilon^{(1)}(k) (\varepsilon^{(2)}(k))$ ranges from 0 to $2w$ ($2v$ 
to $2(w+v)$), and they are separated by a band-gap of width $2(v-w)$, see Fig. \ref{fig1}e. When $w=v$ the two bands merge. In this 
case, the thermalization process leads to a RJ distribution Eq. (\ref{RJ}) as discussed previously (see Figs. \ref{fig1}a-c). 

In the opposite limit of broad gaps ($\Delta\gg1$) Eqs. (\ref{vectors},\ref{SSHspectrum}) can be further simplified to
\begin{eqnarray}
\label{bgexp}
f^{(s)}(k)&\approx&{1\over \sqrt{2}}\left[\cdots,\left(1,(-1)^{s-1}\right),\cdots\right]^T,\\
\varepsilon^{(s)}(k) &\approx& 2v\delta_{s,2} + w \left(1 +(-1)^s \cos(k)\right).\quad s=1,2\nonumber
\end{eqnarray}
It turns out that in this case the system approaches a (quasi-)thermal state which differs from the RJ distribution. In fact, our numerics nicely 
confirms the theoretical predictions of Eq. (\ref{dRJ}) associated with the TCRJ distribution, see Fig. \ref{fig1}f. We found that the 
TCRJ emerges when the ratio of the gap-width to the band-width $\Delta\equiv {v-w\over w}$ reaches the value $\Delta=1$. 

In Fig. \ref{fig1}f we have reported the theoretical predictions (solid red and black lines) for the (quasi-)equilibrium modal-power distribution 
for an SSH system where $\Delta_g=2(v-w)=3/2>\Delta_b=2w=1/2$ where $v=1,w=1/4$. In this broad-gap case we expect the formation of 
TCRJ distribution Eq. (\ref{dRJ}). The optical thermodynamic variables $(T,\mu_1,\mu_2)$ can be evaluated explicitly from the initial beam 
preparation using the extended variables $(E, A^{(1)}, A^{(2)})$, see Eq. (\ref{su1}). We have found that these results are nicely describing the 
outcome of the numerical simulations (see filled circles). We have also confirmed numerically, (see Fig. \ref{fig2}b) that the emergence of the 
TCRJ coincides with the ``individual band power'' conservation of $A^{(1)}, A^{(2)}$.
 
\subsection{Relaxation Times} 

Next, we investigate the relaxation time (distance) that an initial beam needs to propagate before reaching a (quasi)-stationary thermal 
state. Although the analysis below focuses on the case where the broad-gap condition is satisfied, we will also comment on the relaxation 
times for narrow-gap conditions.  
 
We assume a small deviation $\delta I_\alpha$ from the stationary distribution ${\tilde I}_{\alpha}$ such that the modal power of the 
$\alpha$-mode becomes 
\begin{equation}
\label{deviation}
I_\alpha = {\tilde I}_{\alpha} + \delta I_\alpha.
\end{equation}
Substitution of Eq. (\ref{deviation}) in Eq.~(\ref{KE}) and subsequent linearization lead us to the following rate equation 
\begin{equation}
\label{relaxationrate}
\frac{d\delta I_{\alpha}}{dt} \approx -\frac{\delta I_{\alpha}}{\tau_{\alpha}},\quad{\rm where}\quad
\frac{1}{\tau_\alpha} = \frac{\chi^2}{{\tilde I}_{\alpha}}\sum_{\beta\gamma\delta}{\tilde I}_{\beta}{\tilde I}_{\gamma}{\tilde I}_{\delta}
V_{\alpha\beta\gamma\delta} 
\end{equation}
which describes the rate at which a single $\alpha$-mode relaxes towards the thermal state, given that all other modes are already 
at equilibrium. For a better analysis we will decompose the total relaxation rate into various processes associated with 
different collision mechanisms i.e. $\frac{1}{\tau_{\alpha}} = \frac{1}{\tau_{\alpha}^{\rm intra}}+\frac{1}{\tau_{\alpha}^{\rm inter}}$ and 
analyze each one separately. We point out that, in the case of broad gap spectrum, we immediately exclude the inter-band contributions 
associated with the collision process (III). These processes are responsible for a power-exchange between the bands and consequently 
they lead to the formation of a standard RJ thermal state. 

\subsubsection{Intra-band Processes}
We first analyze the intra-band processes (I) that do not contribute to any energy or norm exchange between the bands. These processes 
are also present in case of a narrow-gap (or even single band) photonic system. Below we evaluate the relaxation time of the $\alpha_1-$
modes that belong to the lower band, but a similar calculation can be done for the $\alpha_2-$modes that belong to the upper band. In the 
case of $\alpha_1$-modes, Eq. (\ref{relaxationrate}) becomes
\begin{equation}
\label{intra}
\frac{1}{\tau_{\alpha_1}^{\rm intra}} = \frac{\chi^2}{{\tilde I}_{\alpha_1}}\sum_{\beta_1\gamma_1\delta_1}
{\tilde I}_{\beta_1}{\tilde I}_{\gamma_1}{\tilde I}_{\delta_1}V_{\alpha_1\beta_1\gamma_1\delta_1},
\end{equation}
which can be also derived from the first term of the r.h.s. of Eq. (\ref{normconservation}) after substituting the modal powers $I_{\alpha}$
(see Eq. (\ref{deviation})) and performing a subsequent linearization.

In the case of the SSH model, we can make further progress with the evaluation of Eq. (\ref{intra}). To this end, we substitute Eqs. (\ref{Gamma},
\ref{V-element}) in Eq. (\ref{intra}) and express the matrix $\Gamma_{\alpha_1\beta_1\gamma_1\delta_1}$ in terms of the linear modes of 
the SSH model. When $\Delta\gg 1$, we can use the approximate expressions Eq. (\ref{bgexp}) for the modes of the SSH, which allows
to simplify the matrix $\Gamma_{\alpha_1\beta_1\gamma_1\delta_1}$ into the Kronecker's delta function $\frac{1}{2M}\delta_{k+q-k'-q',0}$, 
where $k,q,k',q'$ refers to the wavenumber of modes $\alpha_1,\beta_1,\gamma_1,\delta_1$ respectively. In the thermodynamic limit 
$M\to\infty$, we turn the sums appearing in Eq. (\ref{intra}) into integrals over the wavenumbers and have that
\begin{align}
\label{intraint}
\frac{1}{\tau^{\rm intra}_k} &\approx&
\frac{\chi^2}{4\pi} \int_{-\pi}^{\pi}dq\int_{-\pi}^{\pi}dk'\frac{{\tilde I}_q}{{\tilde I}_{k}}{\tilde I}_{k'}{\tilde I}_{k+q-k'}\nonumber\\
&\times&\delta(\varepsilon^{(1)}_k+\varepsilon^{(1)}_q-\varepsilon^{(1)}_{k'}-\varepsilon^{(1)}_{k+q-k'}).
\end{align}
The Dirac's delta function in Eq.(\ref{intraint}) enforces four-wave mixing processes which satisfy the 
wavevector constraints $(k,q=\pi-k,k',q'=\pi-k')$. Then Eq.(\ref{intraint}) becomes
\begin{equation}
\label{intraint2}
\frac{1}{\tau^{\rm intra}_k} \approx \frac{\chi^2}{4\pi w}\int_{-\pi}^{\pi}\frac{{\tilde I}_{\pi-k}}{{\tilde I}_{k}}\frac{{\tilde I}_{k'}
{\tilde I}_{\pi- k'}dk'}{\left|\sin k'-\sin k\right|}
\end{equation}
which is amenable to further theoretical treatment in the two limiting cases of high and low temperatures. 

In the low temperature limit $T\rightarrow 0$, the power is concentrated at the bottom of the band corresponding to $\varepsilon^{(1)}
(k\approx 0)\approx 0$ (and similarly $\varepsilon^{(2)}(k\approx \pm \pi)\approx 2v$ for the upper band). The associated modal power 
scales as ${\tilde I}_k\sim {\cal O}(\frac{w}{T})$, while the power of higher modes diminishes as ${\tilde I}_k\sim{\cal O}(\frac{T}{w})$ 
(see Supplementary Material). A substitution of these estimations in Eq. (\ref{intraint2}) leads to the following expression for the 
relaxation times for energies in the middle of the spectrum
\begin{equation}
\label{intra2}
\frac{1}{\tau^{\rm intra}_k} \sim \chi^2 Ta^{(1)}/w^2
\end{equation}
where $a^{(s)}\equiv A^{(s)}/M_s$ is the modal power density associated with the lower $(s=1)$ and upper $(s=2)$ bands respectively.
The scaling of the relaxation times of the modes in the second band is given by a similar relation as Eq. (\ref{intra2}), with the only 
difference being that $a^{(1)}$ has to be substituted with $a^{(2)}$. 

From Eq. (\ref{intra2}) we conclude that the scattering rate vanishes when $T\rightarrow 0$, which is a generic result for any excitation. 
One can also develop a qualitative understanding of the linear dependence of the scattering rate on the temperature $T$ by analyzing 
the expression of the KE appearing in the first line of Eq. (\ref{KE}). For instance, the term  $-I_\alpha I_\beta I_\gamma$, describes the 
following process: an $\alpha$-excitation combines with a $\beta$-excitation and produce an induced emission of a $\gamma$-excitation, 
while the $\delta$-excitation is emitted to conserve energy. This process contributes to the decay rate $1/\tau_\alpha$ for the added 
$\alpha$-excitation a term $\sim I_\beta I_\gamma$. At low $T$, a small energy interval, $\sim T$, is occupied with low energy excitations 
with high density, $\sim 1/T$ (see also Supplementary Material), while the bulk of the band is occupied by ``high-energy'' excitations with 
low density proportional to $\sim T$. Now, for the above process to be possible, the added $\alpha$-excitation (at high energy, in the bulk 
of the band) has to meet a $\beta$-excitation and a $\gamma$-excitation to interact with. Moreover, $\varepsilon_\gamma$ must be close 
to $\varepsilon_\alpha$, in order to conserve energy. Specifically, $\varepsilon_\gamma$ must compensate for the energy of the destroyed 
$\alpha$-excitation in the bulk of the spectrum, while the $\beta$-excitation belongs to the bottom of the band. The large $1/T$ density is 
compensated by the small $T$-factor, coming from the summation over $\beta$, and we are finally left with the $T$-factor due to the small 
density of $\gamma$-excitations so that $1/\tau_\alpha$ is proportional to $T$.

The high temperature limit $T\rightarrow\infty$ of the relaxation times Eq. (\ref{relaxationrate}) can be also evaluated. In this case, 
the power is uniformly distributed within each $\alpha_{s}$-mode in the lower (s=1) and upper (s=2) bands i.e. ${\tilde I}_{\alpha_{s}}=
a^{(s)}$. For the modes in the first band $s=1$ we obtain
\begin{equation}
\label{highTintra}
\frac{1}{\tau_k^{\text{intra}}} \sim \chi^2{(a^{(1)})^2\over w}
\end{equation}
while intra-band relaxation time for the modes in the upper band ($s=2$) will be given by the same expression as Eq. (\ref{highTintra}) 
with the only substitution $a^{(1)}\rightarrow a^{(2)}$. 

The theoretical predictions Eqs. (\ref{intra2},\ref{highTintra}) are nicely confirmed by a direct numerical evaluation of the intra-band relaxation 
times using Eq. (\ref{intra}), see Fig. \ref{fig3}a. Let us also mention that the intra-band transitions discussed above have their equivalent 
in narrow gap systems as well (for the high temperature limit see Ref. \cite{B14}). In this case, of course, one has a standard RJ distribution 
Eq. (\ref{RJ}) of the modal powers. As the power 
populates progressively the modes of the lower band for $T\rightarrow 0$, we have to substitute in Eq. (\ref{intra2}), the individual band power 
density $a^{(1)}$ with the total one $a\equiv A/M$. For completeness, we also report in Fig. \ref{fig3}b, the scaling behavior of the intra-band 
relaxation times for the narrow-gap scenario of an SSH model corresponding to inter-dimer coupling $w=2/3$ and intra-dimer $v=1$.

\subsubsection{Inter-band Processes}

Next we analyze the contribution of collision processes (II) to the relaxation of the $\alpha_1$-modes. This mechanism involves inter
-band transitions and it is responsible for the energy relaxation between the two bands. In this case, Eq. (\ref{relaxationrate}) becomes
\begin{equation}
\label{inter}
\frac{1}{\tau_{\alpha_1}^{\rm{inter}}} = 2\frac{\chi^2}{{\tilde I}_{\alpha_1}}\sum_{\beta_2\gamma_1\delta_2}{\tilde I}_{\beta_2}
{\tilde I}_{\gamma_1}{\tilde I}_{\delta_2}V_{\alpha_1\beta_2\gamma_1\delta_2} 
\end{equation}
which can be evaluated analytically in the case of the SSH model. Following the same calculations as for the intra-band relaxation time, 
we can further simplify the above equation into the following form
\begin{eqnarray}\label{SSHinter}
\frac{1}{\tau_{k}^{\rm{inter}}} \approx \frac{\chi^2}{2\pi w}\int_{-\pi}^{\pi}\frac{{\tilde I}_{-k}^{(2)}}{{\tilde I}_k^{(1)}}\frac{{\tilde I}_{-k'}^{(1)}
{\tilde I}_{k'}^{(2)}dk'}{\left|\sin k'-\sin k\right|},
\end{eqnarray}
where the super-indexes in the modal powers indicate the corresponding band. In the above analysis, we have used the fact that the 
presence of the Kronecker's and Dirac's delta functions (see Eqs. (\ref{Gamma},\ref{V-element}), and the discussion in the previous 
subsection) limits the four-wave mixing to processes that satisfy the relations $(k,q=-k,k',q'=-k')$ . 

When $T\to\infty$, the modal power becomes independent of $k$ and it only depends on the power density associated with the specific 
band $s$, i.e. ${\tilde I}_k^{(s)}=a^{(s)}$. Using Eq. (\ref{SSHinter}) we get that the inter-band relaxation rate of the modes 
associated with the first band is
\begin{equation}
\label{highTinter}
\frac{1}{\tau_{k}^{\rm{inter}}} \sim \frac{\chi^2}{w}(a^{(2)})^2 
\end{equation}
A similar expression applies for the modes of the second band with the obvious substitution $a^{(2)}\rightarrow a^{(1)}$.

In the opposite limit of $T\to 0$ the power will be mainly concentrated in modes that are at the bottom of the spectrum of each of the 
two bands. Specifically, the modal power ${\tilde I}_k^{(1)}$ at the first band will be significant for modes with $k\approx 0$, while the 
modal power ${\tilde I}_k^{(2)}$ at the second band will be appreciable for modes with $k\approx \pm \pi$. Using their corresponding 
approximate expressions (see Supplementary Material) and substituting them back in Eq. (\ref{SSHinter}) we have
\begin{equation}
\label{lowTinter} 
\frac{1}{\tau_{k}^{\rm{inter}}}\sim \frac{\chi^2 T}{w^2}{[a^{(1)}+a^{(2)}]\over 2}
\end{equation}
which holds for modes belonging to either band. The relaxation rates vanish linearly with the temperature $T$ as in the intra-band 
scenario discussed above. In fact, one can invoke the same considerations as above, for the qualitative understanding of this linear 
scaling law. 

\begin{figure}
\centering
\includegraphics[width=1\columnwidth]{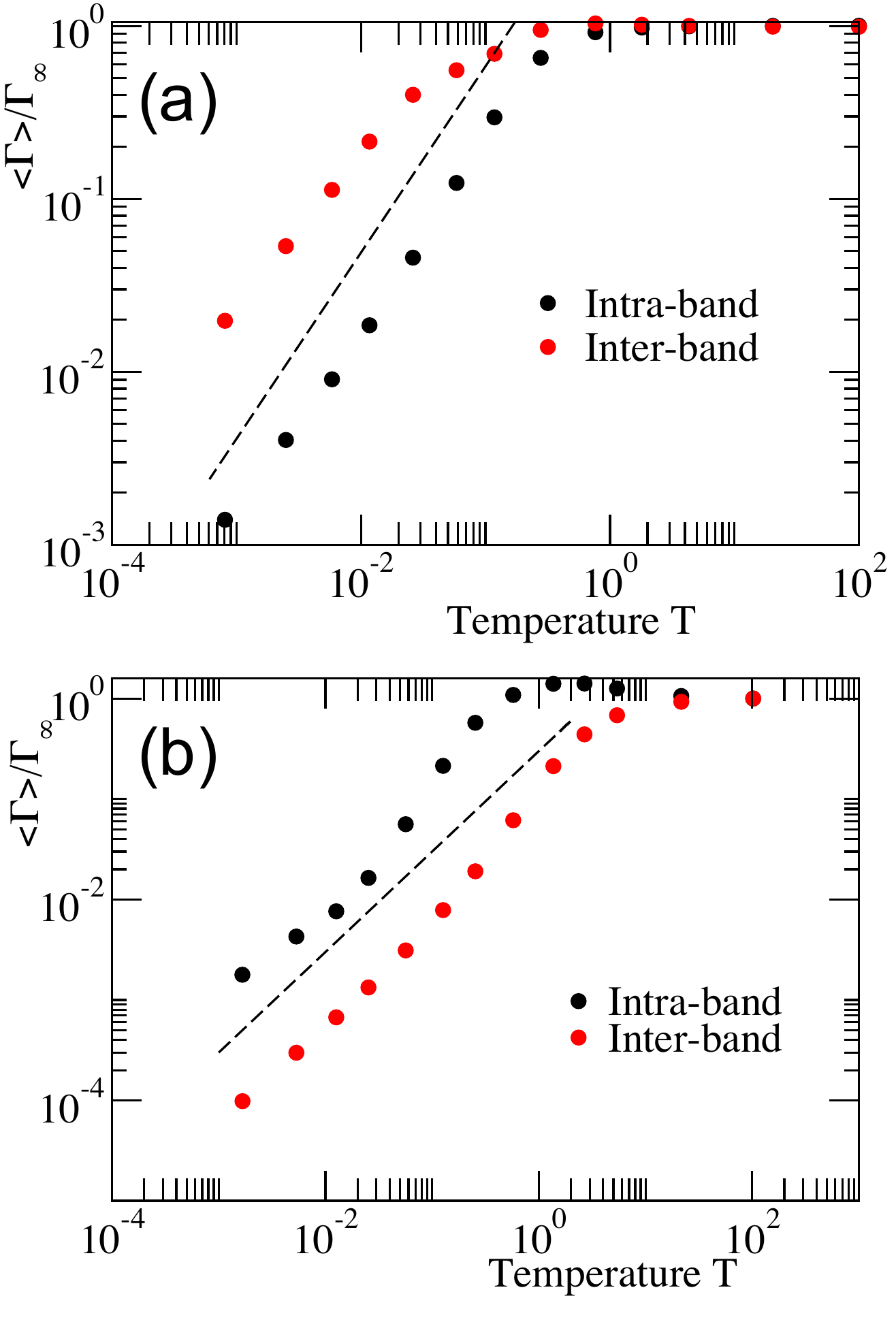}
\caption{Scaling of relaxation rates $\Gamma_k\equiv1/\tau_k$ versus temperature for modes in the middle of the lower band: (a) the broad-gap 
scenario associated with an SSH model (see Fig. \ref{fig1}d) with inter-dimer coupling $w=1/3$ and intra-dimer $v=1$; (b) the narrow-gap 
scenario for an SSH model with inter-dimer coupling $w=2/3$ and intra-dimer $v=1$. The numerical data have been extracted via an evaluation 
of Eq. (\ref{relaxationrate}) for the SSH model with $M=32$. The black dashed line has slope one and it is drawn in order to guide the eye. 
}
\label{fig3}
\end{figure}

The above theoretical predictions for the inter-band relaxation times describe nicely our results from the direct numerical evaluation of 
Eq. (\ref{inter}) for the SSH model in case where the broad gap condition Eq. (\ref{ab3}) is satisfied, see Fig. \ref{fig3}a. In the opposite 
case of narrow-gaps (but still $\Delta\neq 0$), the inter-band relaxation rate is supplemented with an additional term associated with 
collision processes (III). Although we were not able to evaluate these additional contributions analytically, our detailed numerical calculations 
using Eq. (\ref{relaxationrate}) confirm that these processes do not affect the overall behavior of the relaxation rates, see Fig. \ref{fig3}b. 

\subsubsection{Limitations and Energy Relaxation Rate}

The main assumption underlying the derivation of the KE is that initially the phases of the linear modes are random and only the intensities 
are out of equilibrium.  However, for an arbitrary initial preparation the phases need not be random and, generally, some ``phase randomization 
time'', $\tau^{\prime}$, may enter the picture. For small $J_{l,j}$, this time-scale, is controlled by the very slow process of equilibration between 
different regions in real space and it obviously approaches 
infinity when $J_{l,j}\rightarrow 0$. In typical circumstances where the coupling between local modes is evanescently small, we expect that 
$\tau^{\prime}\geq \tau_{\alpha}$ and therefore the latter can be always considered as the lower bound of the total relaxation process. 

Let us finally point out that our analysis also sheds light on energy (not power!) relaxation between the two bands. For example, the rate of 
energy exchange of the first band $\delta E^{(1)}\equiv\sum_{\alpha'}\varepsilon_{\alpha_1}^{(1)}
\delta I_{\alpha_1}$ can be approximated as
\begin{equation}\label{energyrelaxation}
\frac{d\delta E^{(1)}}{dt} \approx \sum_{\alpha_1}\varepsilon_{\alpha_1}^{(1)}\frac{d\delta I_{\alpha_1}}{dt} \approx -\sum_{\alpha_1}
\varepsilon_{\alpha_1}^{(1)}\frac{\delta I_{\alpha_1}}{\tau_{\alpha_1}^{\rm{inter}}}
\end{equation}
where we have used Eq. (\ref{relaxationrate}) together with the fact that only the inter-band transitions can contribute to the energy relaxation 
between bands. Equation~(\ref{energyrelaxation}) can be used for extracting the energy relaxation time $\tau_{H_1}$ defined as {$\frac{d 
\delta E^{(1)}}{dt} \equiv-{\delta E^{(1)}\over \tau_{H}}$. 

\section{Thermalization Management in Random Photonic Networks}\label{random} 

In the previous section we have highlighted the importance of dispersion engineering in the control of the thermalization process of an initial 
beam. Our analysis was focusing on photonic networks with an underlying spatial periodicity. It is natural, therefore, to question the validity of 
our theoretical results in cases of random photonic networks where the concepts of bands and gaps are not, strictly speaking, applicable. Still, 
one can design photonic structures whose spectrum is consisting of groups of modes which are spectrally away from one another and investigate 
the applicability of the ``broad gap'' condition for the realization of TCRJ distributions. In this section we are analyzing the thermal state for two 
such representative non-periodic networks. 

The first network consists of an array of coupled single-mode waveguides (resonators). The propagation constants (node-resonant frequencies) 
are $J_{ll}=J_1$ for the first $n=1,\cdots,M_1$ waveguides (resonators) and $J_{ll}=J_2$ for the remaining $n=1,\cdots, M_2$ waveguides 
(resonators) \cite{KS11}. For simplicity we have further assumed that $M_1=M_2$ while we considered that all waveguides (resonators) 
are coupled together with the same coupling constant $J_{n,n\pm 1}=J_0$. The parameters $(J_0, J_1, J_2)$ of the network are chosen 
in a way that the spectrum is organized in two groups that are separated by a gap which satisfies a broad gap condition Eq. (\ref{ab3}), see Fig. 
\ref{fig4}a. In Fig. \ref{fig4}b we show the (quasi-)thermal state (filled circles) associated with a specific initial excitation (open circles). We have 
extracted the optical temperature $T$ and the two chemical potentials $(\mu_1,\mu_2)$ from the individual-band powers $A^{(1)}, A^{(2)}$ 
and the total energy $H$ of the initial excitation, see Eqs. (\ref{su1}). Using these inputs, we have evaluated the TCRJ distribution Eq. (\ref{dRJ}) 
for the modal powers (see red and black lines in Fig. \ref{fig4}b). The theoretical results, nicely match the numerical data extracted by evolving 
the initial excitation up to times $t\approx 10^7$ (units of $J_0$).

A similar analysis has been performed for a random photonic network whose connectivity matrix $J_{nm}$ has been constructed via an 
orthogonal transformation of a diagonal matrix $\Lambda_{nm}=\delta_{nm} \left[nJ_0+(\Delta_g-J_0)\Theta(n-M_1)\right]$ i.e., $J=O^T
\Lambda O$ where $O$ is a random orthogonal matrix. The linear spectrum of this 
network is characterized by the parameters $J_0, M_1$, and $\Delta_g$ which have been chosen to enforce a broad-gap (see Eq. 
(\ref{ab3})) between two spectral groups, see Fig. \ref{fig4}c. Following the same methodology as previously, we have extracted from Eqs.
(\ref{su1}) the theoretical values for the optical temperature $T$ and the chemical potentials $(\mu_1,\mu_2)$ using as input the individual-
band powers $A^{(1)}, A^{(2)}$ and the total energy $E$ of the initial excitation. The predicted TCRJ, which uses these values, is shown in 
Fig. \ref{fig4}d together with the modal thermal occupation that has been evaluated via direct dynamical simulations. In these simulations 
the initial excitation was evolved up to time $t\approx 10^5$. The nice comparison confirms once more the validity of our theory. 

\section{Conclusions}\label{conclusions}

We have highlighted the importance of dispersion engineering methods for the control of the thermalization process and the formation of 
a thermal state of an initial beam propagating in a nonlinear multimode photonic structure. Using a kinetic equation  
approach we have shown that its stationary solutions might differ from the standard Rayleigh-Jeans distribution if the ratio between 
the gap-width and the band-width exceeds a critical value. In such case, each individual band preserves the power with which it is initially
populated. The modal power in each band is characterized by a Rayleigh-Jeans distribution with distinct chemical potential which is dictated by the 
individual band power. The latter, together with the optical temperature that it is determined by the initial beam, controls the relaxation 
rate towards these thermal states. We have tested numerically the validity of the predictions of the kinetic equation using the SSH model whose band-
gap structure is controlled by the coupling between the elements of the network. We have further extended the spectral engineering rules 
for the management of the thermal states using more complex networks that are not periodic. We have shown that an appropriate connectivity 
between the elements of the network, can enforce the formation of groups of modes that are separated by spectral gaps whose width determines 
the thermal state of 
the system. It will be interesting to implement these ideas to other complex systems, like quasi-periodic or aperiodic photonic structures 
that demonstrate fractal spectra, and analyze the thermalization process towards a thermal equilibrium state.

\begin{figure}
\centering
\includegraphics[width=1\columnwidth]{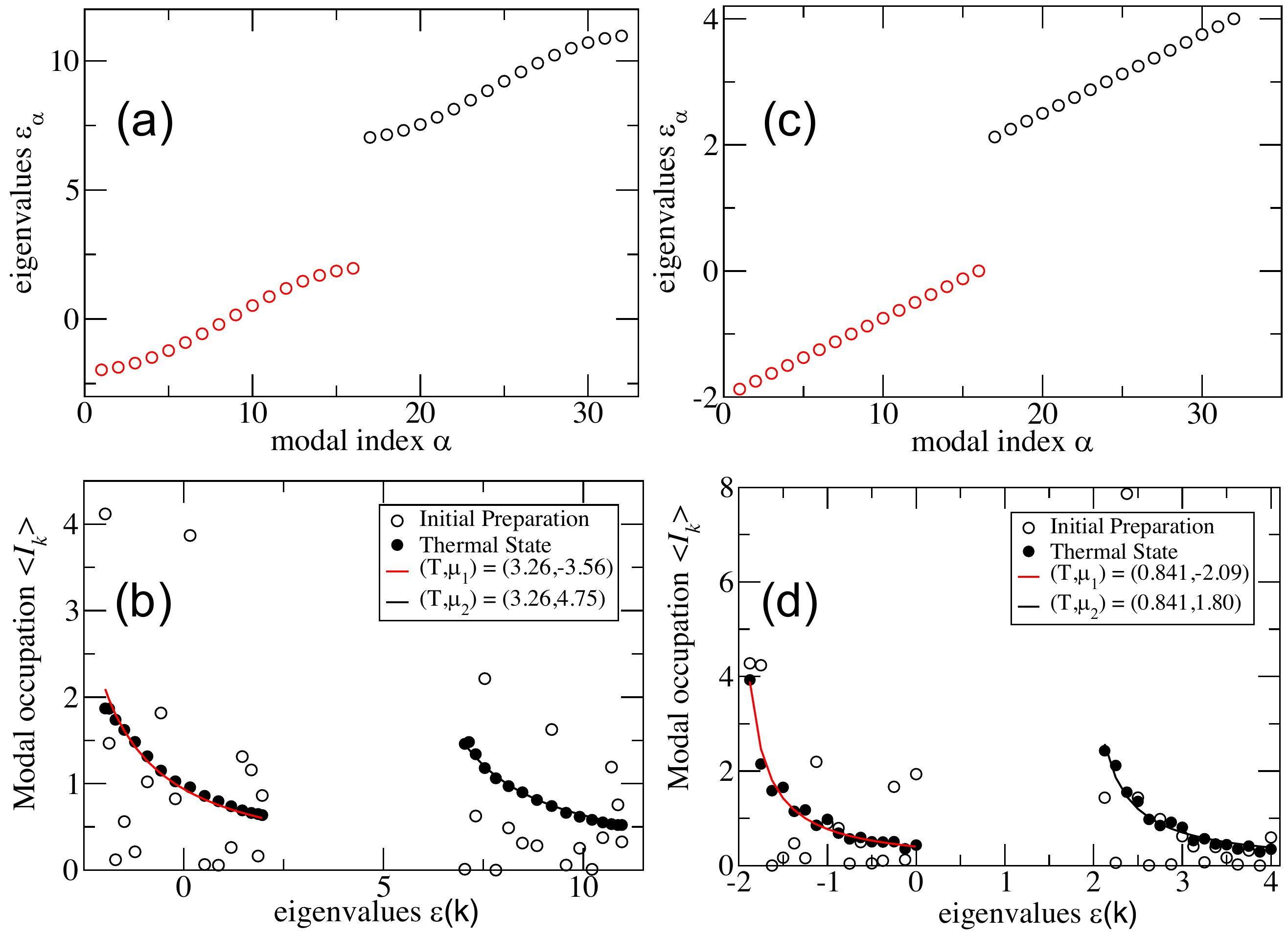}
\caption{(a)Spectrum $\varepsilon_{\alpha}$ vs. modal index $\alpha=1,\cdots, M$ of a photonic circuit consisting of two arrays of $M=32$ 
nonlinear waveguides (resonators) coupled together with a constant coupling strength $J_{0}=1$. The first $M_1=16$ single-mode 
waveguides have the same propagation constant $J_{1}=0$, while the remaining $M_2=16$ waveguides have another propagation constant 
$J_{2}=9$. The two propagation constants have been chosen in a way that the spectrum satisfies a broad-gap condition $\Delta=1.25$ 
corresponding to $\Delta_g=5$ and $\Delta_b=4$; (b) The modal power distribution of the initial excitation (open circles) and of the thermal 
state (filled black circles). The latter is extracted from the temporal evolution of the initial preparation. The red and black solid lines indicate 
the TCRJ distribution with $(T,\mu_1,\mu_2)=(3.26,-3.56,4.75)$. The Kerr nonlinearity parameter used in this simulation is $\chi=0.1$; (c) The 
spectrum of a network of $M=32$ randomly connected coupled resonators. The network is described by a connectivity matrix matrix $J_{lj}$ 
(see details in the text) with spectrum consisting of equally-spaced eigenmodes that populate two intervals $[-1.875,0]$ and $[2.125,4]$. The 
gap-width between these two spectral intervals is $\Delta_g=2.125$ corresponding to broad-gap condition $\Delta=1.1333$; (d) The same as 
in (b) but for modal power distribution associated with the random network of (c). The parameters that determine the TCRJ distribution in this 
case are $(T,\mu_1,\mu_2)=(0.841,-2.09,1.8)$. The Kerr nonlinearity parameter used in this simulation is $\chi=0.01$. 
}
\label{fig4}
\end{figure}

\emph{Acknowledgments -- } 
C.S. and T.K. acknowledge partial support from grant ONR N00014-16-1-2803, and from a grant from the Simons Foundation for Collaboration 
in MPS No. 733698. C.S. acknowledges Dr. A. Ramos for technical support with the symplectic code for the dynamical simulations.



\clearpage

\section{Supplementary Material}
\beginsupplement

\setcounter{section}{0}
\setcounter{equation}{0}
\renewcommand{\thesection}{S\arabic{section}}  
\renewcommand{\theequation}{S\arabic{equation}}
\section{Low temperature approximation of modal power in a standard RJ distribution}

In this section we derive the relaxation times for the one-band spectrum. In this case the dispersion relation of the system is $\varepsilon_{\alpha}
\equiv\varepsilon(k)=-2w\cos(k)$ where the wavevector $k\in[-\pi,\pi]$ and $w$ is the coupling element between nearby waveguides (or resonators). 
Our starting point is the general equation Eq. (19) that give us the relaxation times 
$\tau_{\alpha}$ as:
\begin{equation}
\label{1band1}
\frac{1}{\tau_\alpha} = \frac{4\pi\chi^2}{{\tilde I}_{\alpha}}\sum_{\beta\gamma\delta}{\tilde I}_{\beta}{\tilde I}_{\gamma}{\tilde I}_{\delta}
\Gamma_{\alpha\beta\gamma\delta}^2 \delta(\varepsilon_{\alpha}+
\varepsilon_{\beta}-\varepsilon_{\gamma}-\varepsilon_{\delta})
\end{equation}
where $\Gamma_{\alpha\beta\gamma\delta}$ is given by Eq. (5). In the large system-size $M\rightarrow \infty$ limit the above 
sum can be turned to the following integral
\begin{widetext}
\begin{eqnarray}
\label{1band2}
\frac{1}{\tau_k} &=&\frac{\chi^2}{\pi{\tilde I}_{k}}\int_{-\pi}^{\pi}\int_{-\pi}^{\pi}
{\tilde I}_{q}{\tilde I}_{q'}{\tilde I}_{k+q-q'}\delta\left(-2w\left[\cos(k)+\cos(q)-\cos(q')-\cos(k+q-q')\right]\right)dqdq'\nonumber\\
&=&\frac{\chi^2}{2\pi w{\tilde I}_{k}}\int_{-\pi}^{\pi}\frac{dq'}{\left|\sin(q')-\sin(k)\right|}{\tilde I}_{\pi-k}{\tilde I}_{q'}{\tilde I}_{\pi-q'}
\end{eqnarray}
\end{widetext}
In the infinite temperatire limit $T\rightarrow \infty$, the equilibrium distribution of the modal powers is ${\tilde I}_k={-T\over \mu}=a\equiv 
A/M$ ($\mu<0$) and the above expression becomes
\begin{equation}
\label{1band3}
\frac{1}{\tau_k}=\frac{\chi^2}{2\pi w}a^2\int_{-\pi}^{\pi}\frac{dq'}{\left|\sin(q')-\sin(k)\right|}.
\end{equation}
The above expression has been also derived by Basko in Ref. \cite{B14}.

The analysis of the low temperature limit $T\rightarrow 0$ of the relaxation times is more subtle. To this end we first express the 
temperature and the chemical potential in terms of the external variables $(E,A)$ that characterized the initial beam. Specifically 
\begin{eqnarray}
\label{1band4}
\sum_{\alpha = 1}^{M}\frac{T}{\varepsilon_{\alpha}-\mu} &=& {-M\over 2\pi}\int_{-\pi}^{\pi} \frac{T}{2w\cos(k)+\mu}dk=\frac{MT}
{\sqrt{\mu^2-4w^2}}=A,\nonumber\\  
\sum_{\alpha = 1}^{M}\frac{\varepsilon_{\alpha}T}{\varepsilon_{\alpha}-\mu} &=&M T + \mu A= E
\end{eqnarray}
From the above equations we find that 
\begin{equation}
\label{1band5}
T=\frac{h^2-4w^2a^2}{2h};\quad \mu=\frac{h^2+4w^2 a^2}{2ah}
\end{equation}

Next, recall that in the Gibbsian region both $h$ and $\mu$ are negative. Specifically, for zero temperature $h=-2wa$ while for
infinite temperature we have $h\rightarrow 0$. Similarly, $\mu$ changes from $-2w$ to $-\infty$. Having this in mind, we write for 
low temperatures $T\rightarrow 0$, the energy density $h$ as $h\approx  -2wa +\eta$ with $\eta\ll 1$. It follows from Eq. (\ref{1band5}) 
that 
\begin{equation}
\label{1band6}
T\approx \eta;\quad \mu\approx -2w-{\eta^2\over 4w a^2}
\end{equation}
which allows us to express the modal powers appearing in Eq. (\ref{1band2}) as
\begin{equation}
\label{1band7}
{\tilde I}_{q}=\frac{\eta}{2wa+{\eta^2\over4wa^2}-2w\cos(q)}.
\end{equation}
At low temperatures $T$, the power is concentrated in a narrow interval $\delta q \sim {T\over wa}$ near the bottom of the band. 
In that interval ${\tilde I}_q \sim w {a^2\over T}$, and everywhere else in the band ${\tilde I}_k \sim {T\over w}$. Assuming a generic 
mode $k$ in the bulk of the band, the factor ${\tilde I}_{\pi-k}/ {\tilde I}_k\sim {\cal O}(1)$. Furthermore, the main contribution to the 
integral in Eq. (\ref{1band2}) comes from the narrow interval $\delta q$ where ${\tilde I}_{q'}\times {\tilde I}_{\pi-q'}\sim w{a^2\over T}
\times {T\over w}$ so that
\begin{equation}
\label{1band8}
1/\tau_k \sim {\chi^2\over w}{wa^2\over T}{T\over w} \delta q' = {\chi^2\over w} a^2 {T\over wa} =\chi^2 a {T\over w^2}
\end{equation}   
which demonstrates the same scaling behavior with $T$, as the one occuring for intraband transitions Eq. (23) and for interband transitions 
Eq. (28) in the low temperature limit. The above estimations ignore a logarithmic factor which appears after regularization of 
the integral appearing in Eq. (\ref{1band2}).

\section{Low temperature approximation of modal power in a GRJ distribution}

The total energy and norm in each band is 
\begin{eqnarray}
\label{Ssub1}
\sum_{\alpha_1 = 1}^{M_1}\frac{T}{\varepsilon_{\alpha_1}^{(1)}-\mu_1} &=& A^{(1)},\indent \sum_{\alpha_2 = 1}^{M_2}\frac{T}
{\varepsilon_{\alpha_2}^{(2)}-\mu_2} = A^{(2)}, \\
E&=&(M_1+M_2) T + \mu_1 A^{(1)} + \mu_2 A^{(2)}\nonumber
\end{eqnarray}
where in the case of the SSH model $M_1=M_2$ is the number of modes in band $s=1$ and $s=2$ respectively and $M=2M_1$
is the total numger of modes in both bands.
 
Furthermore, we define $h\equiv {\frac{E}{M}}$ to be the energy density, and $a^{(1)} \equiv {\frac{A^{(1)}}{M_1}}$, $a^{(2)}\equiv
{\frac{A^{(2)}}{M_2}}$ to be the individual band power density. Under the broad-gap approximation $\Delta \gg 1$, we can substitute 
from Eq. (17) the eigenvalues of the SSH model, and arrive to the following expressions 
\begin{gather}
\label{sub2}
\frac{1}{2\pi}\int_{-\pi}^{\pi}\frac{T\cdot dk}{w(1-\cos k)-\mu_1} = a^{(1)}, \nonumber\\
\frac{1}{2\pi}\int_{-\pi}^{\pi}\frac{T\cdot dk}{(2v+w(1+\cos k))-\mu_2} = a^{(2)}, \nonumber\\
T + \frac{\mu_1 a^{(1)}}{2} + \frac{\mu_2 a^{(2)}}{2} = h.
\end{gather}
The two integrals above can be evaluated analytically, thus allowing us to express Eq. (\ref{Ssub1}) as
\begin{eqnarray}
\label{sub3}
a^{(1)}=\frac{T}{\sqrt{(\mu_1 - w)^2 - w^2}} ,&
a^{(2)}=\frac{T}{\sqrt{(\mu_2 - w - 2v)^2 - w^2}} , \nonumber\\
h=T + \frac{\mu_1 a^{(1)}}{2} + \frac{\mu_2 a^{(2)}}{2}. &
\end{eqnarray}
We proceed with the evaluation of $(T,\mu_1,\mu_2)$ from the above set of three algebraic equations. To this end we define
$h^{(1)}\equiv {1\over M_1}\sum_{\alpha_1 = 1}^{M_1}\frac{\varepsilon_{\alpha_1}^{(1)} T}{\varepsilon_{\alpha_1}^{(1)}-\mu_1}$
and similarly $h^{(2)}\equiv {1\over M_2}\sum_{\alpha_2 = 1}^{M_2}\frac{\varepsilon_{\alpha_2}^{(2)} T}{\varepsilon_{\alpha_2}^{(2)}
-\mu_2}$ such that $h^{(1)}+h^{(2)}=h$. Using these new variables we are able to rewrite Eq. (\ref{sub3}) in the following form
\begin{eqnarray}
\label{sub4}
a^{(1)}=\frac{T}{\sqrt{(\mu_1 - w)^2 - w^2}},&  h^{(1)}=T + \mu_1 a^{(1)}\quad\quad\\
a^{(2)}=\frac{T}{\sqrt{(\mu_2 - w - 2v)^2 - w^2}}, & h^{(2)}=T + \mu_2 a^{(2)}. \quad\nonumber
\end{eqnarray}
From the first set of equations we get $T = h^{(1)}\frac{h^{(1)}-2wa^{(1)}}{2(h^{(1)}-wa^{(1)})}$ and $\mu_1 = \frac{{h^{(1)}}^2}
{2a^{(1)}(h^{(1)}-wa^{(1)})}$ and similarly $T=\frac{(h^{(2)}-2a^{(2)}v)^2-2a^{(2)}w(h^{(2)}-2a^{(2)}v)}{2(h^{(2)}-2a^{(2)}v-a^{(2)}w)}$, 
and $\mu_2 = 2v + \frac{(h^{(2)}-2a^{(2)}v)^2}{2(a^{(2)}h^{(2)}-w{a^{(2)}}^2-2v{a^{(2)}}^2)}$. Assuming that $h^{(1)}\ll w$, the first set 
of equations can be simplify further and give us $T\approx h^{(1)}+\mathcal{O}({h^{(1)}}^2)$ and $\mu_1\approx -\frac{{h^{(1)}}^2}
{2w{a^{(1)}}^2}+\mathcal{O}({h^{(1)}}^3)$. From the second set of equations associated with the second band, we derive similar 
results i.e. $T\approx\eta+\mathcal{O}(\eta^2)$ and $\mu_2\approx 2v-\frac{\eta^2}{2w{a^{(2)}}^2}+\mathcal{O}(\eta^3)$ when $\eta 
\equiv h^{(2)}
-2va^{(2)}\ll w$. The variable $\eta$ in these expressions indicates the difference of the energy density of the second band, when the 
thermal state has been reached, from its lowest accessible value.

Substituting these expressions for the optical chemical potentials in the modal power occupations for the GRJ we get: 
\begin{equation}
    \Tilde{I}^{(1)}_k \approx \frac{T}{w(1-\cos k)+\frac{T^2}{2w{a^{(1)}}^2}},
    \Tilde{I}^{(2)}_k \approx \frac{T}{w(1+\cos k)+\frac{T^2}{2w{a^{(2)}}^2}}.
\end{equation}
From here, using the same line of argumentation as in the one-band case, we derive the following expressions for the modal powers 
at the first band
\begin{equation}
\label{modal1}
I^{(1)}_{k}\sim\left\{
\begin{array}{ccc}
\left(\frac{w}{T}\right)[a^{(1)}]^2 & {\rm for} & k\ll\frac{T}{w\cdot a^{(1)}}\\[1mm]
\left(\frac{T}{w}\right) & {\rm for} & k\gg\frac{T}{w\cdot a^{(1)}}
\end{array}
\right.
\end{equation}
Similarly, for the modal powers at the second band we have
\begin{equation}
\label{modal2}
I^{(2)}_{k}\sim\left\{
\begin{array}{ccc}
(\frac{w}{T})[a^{(2)}]^2 & {\rm for} & |k-\pi|\ll\frac{T}{w\cdot a^{(2)}}\\[1mm]
(\frac{T}{w}) & {\rm for} & |k-\pi|\gg\frac{T}{w\cdot a^{(2)}}
\end{array}
\right.
\end{equation}

\section{Parameter domain for Gibbsian Thermalization}

Thermalization towards a standard RJ with positive temperature occurs in a certain region of the $(h,a)$ plane where $h\equiv E/M$ is 
the energy density and $a\equiv A/M$ is the power density. Within this range of parameter space, the system thermalizes in accordance 
to the Gibbsian formalism. We point out, however, that one can also achieve thermalization with negative temperatures in case of finite 
number of modes $M$. In this latter case, the higher-order modes are mostly occupied while thermalization with positive temperatures 
indicates a scenario where the lower-energy group of modes is occupied. This situation is consistent  with the beam self-cleaning phenomenon. 
The borders of the Gibbsian domain (for a fixed average power $a$) are defined by the maximum energy density (upper border) corresponding 
to $T\rightarrow \infty$ and by the minimum energy density (lower border) corresponding to $T\rightarrow 0$. It turns out that these two limits 
are amenable to analytical treatment for the case of a periodic photonic networks. For our analysis below we will be using the normal 
mode representation of the Hamiltonian 
\begin{equation}
\label{CMTC}
{\cal H}(\{C_{\alpha}\})=\sum_{\alpha=1}^M \varepsilon_{\alpha} |C_{\alpha}|^2 +{\chi\over 2}\sum_{\alpha\beta\gamma\delta}
\Gamma_{\alpha\beta\gamma\delta}C_{\alpha}^*C_{\beta}^*C_{\gamma}C_{\delta}=E
\end{equation}
with the coupling matrix $\Gamma_{\alpha\beta\gamma\delta}$ given by Eq. (5). We will also assume that the spectrum of 
the linear system is ordered as $\epsilon_1<\epsilon_2<\cdots<\epsilon_M$.

The total power ${\cal N}$ in the normal mode basis is
\begin{equation}
\label{CMTN}
{\cal N}(\{C_{\alpha}\})=\sum_{\alpha=1}^M |C_{\alpha}|^2=A
\end{equation}

\subsection{One band systems}
 
First we analyze the $(h,a)$ space diagram corresponding to a periodic system with coupling constants $J_{i,j}=J_0$ (see Eq. (2)). 
In this case the normal modes are $f_{\alpha}(l)={1\over\sqrt{M}}e^{ik_{\alpha}l}$ ($k_{\alpha}$ is the wavevector associated with 
the $\alpha-$th mode) resulting in a coupling matrix 
\begin{equation}
\label{Gmatrix}
\Gamma_{\alpha\beta\gamma\delta}={1\over M^2}\sum_l e^{i(k_{\alpha}+k_{\beta}-k_{\gamma}-q_{\delta})l}={1\over M}
\delta_{k_{\alpha}+q_{\beta}-k_{\beta}-q_{\delta},0}
\end{equation}
where $\delta_{n,l}$ is the Kroneker delta.

In the low temperature limit $T=0$ only the ground state mode, $C_1$ is occupied. The ground state modal power is evaluated from Eq. 
(\ref{CMTN}) and takes the value $|C_1|^2=Ma$. Substituting this expression together with Eq. (\ref{Gmatrix}) in Eq. (\ref{CMTC}) 
we get the following expression for the minimum energy density 
\begin{equation}
\label{minh}
h_{\rm min}={1\over M}\left(\epsilon_1 a M+{1\over 2}\chi |C_1|^4\Gamma_{1,1,1,1}\right)=\epsilon_1 a +{1\over 2}\chi a^2.
\end{equation}

In the opposite limit of high temperatures $T\rightarrow \infty$ all modes are equally excited. From Eq. (\ref{CMTN}) we get  
$C_{\alpha}=\sqrt{a} e^{i\Phi_{\alpha}}$ where $\Phi_{\alpha}$ is a random phase. Substituting the expression for the occupation 
amplitudes $C_{\alpha}$ in Eq. (\ref{CMTC}) we get the following expression for the maximum energy density
\begin{widetext}
\begin{equation}
\label{maxh}
h_{\rm max}={1\over M}\left(a\sum_{\alpha}\varepsilon_{\alpha}+{\chi\over 2}\sum_{\alpha\beta\gamma\delta}\Gamma_{\alpha\beta\gamma\delta}
\langle C^*_{\alpha}C^*_{\beta}C_{\gamma}C_{\delta}\rangle\right)=
a\overline{\mathcal{E}}+\chi a^2
\end{equation}
\end{widetext}
where $\overline{\mathcal{E}}\equiv {1\over M}\sum_{\alpha}\varepsilon_{\alpha}$ is the mean energy of the spectrum and $\langle\cdots
\rangle$ indicates an averaging over random phases. The last step in the equation above utilized the following contraction rule
\begin{widetext}
\begin{equation}
 \langle C^*_{\alpha}C^*_{\beta}C_{\gamma}C_{\delta}\rangle=\langle C_{\alpha}^*C_{\gamma}\rangle\langle C_{\beta}^*C_{\delta}\rangle+
\langle C_{\alpha}^*C_{\delta}\rangle\langle C_{\gamma}^*C_{\beta}\rangle=I_{\alpha}I_{\beta}\left(\delta_{\alpha\gamma}\delta_{\beta\delta}
+\delta_{\alpha\delta}\delta_{\beta\gamma}\right)
\end{equation}
\end{widetext}

\subsection{Two-band systems}
We proceed with a similar analysis for a two-band system. We assume that the number of modes in the lower band is $M_1$ and in the 
upper band is $M_2$. The total number of modes is $M=M_1+M_2$. In case of GRJ thermal states, one needs to take into consideration 
the additional ``individual-band'' conservation laws:
\begin{equation}
\label{lcl}
A_1=\sum_{\alpha_1}|C_{\alpha_1}|^2=M_1a^{(1)};\quad A_2=\sum_{\alpha_2}|C_{\alpha_2}|^2 =M_2a^{(2)}
\end{equation}
where $a^{(s)}$ are the power densities in each band $s=1,2$

The low temperature limit $T=0$ is trivial. Following the same arguments as for the one-band case, we get for the minimum energy density 
(for fixed $a_1,a_2$):
\begin{equation}
\label{2hmin}
h_{\rm min}=\sum_{s=1,2}\left(\epsilon_1^{(s)} a^{(s)} m_s+{1\over 2}\chi (m_s a^{(s)})^2\right)
\end{equation}
where we have defined $m_s\equiv M_s/M$.

The evaluation of the high temperature border is more subtle. Before writing it down, we note that only $\Gamma_{\alpha\beta
\alpha\beta}=\Gamma_{\alpha\beta\beta\alpha}$ come into play and regardless of whether $\alpha,\beta$ belong to the same band or 
to different bands we have
\begin{equation}
\label{2gamma}
\Gamma_{\alpha\beta\alpha\beta}=\Gamma_{\alpha\beta\beta\alpha}={1\over M}
\end{equation}
Following the same argumentation as in the case of one band we assume that all modes are equally excited. From Eq. (\ref{lcl}) we 
conclude that $C_{\alpha_s}=\sqrt{a^{(s)}}e^{i\Phi_{\alpha_s}}$. It then immediately follows that the total maximum energy density is
\begin{equation}
\label{2hmax}
h_{\rm max}=\sum_sm_sa^{(s)}\overline{\mathcal{E}^{(s)}}+\chi\left(m_1a^{(1)}+m_2a^{(2)}\right)^2
\end{equation}


\end{document}